# Role of Ion Milling Angle in Determining Conducting and Insulating States on SrTiO$_3$ Surfaces

Yuki K. Wakabayashi[a)], Yoshiharu Krockenberger, Kosuke Takiguchi, Hideki Yamamoto, and Yoshitaka Taniyasu

*NTT Basic Research Laboratories, NTT Corporation, Atsugi-shi, Kanagawa 243-0198, Japan*

a)Author to whom correspondence should be addressed: yuuki.wakabayashi@ntt.com

**ABSTRACT**

SrTiO$_3$ (STO), a promising wide-bandgap semiconductor for high-$k$ capacitors and photocatalysis, requires precise surface control for device fabrication. This study investigates the impact of ion milling on STO's surface conductivity. We find that ion milling at incident angles below 10° preserves the insulating state, while ion milling at larger angles induces a conducting surface with high electron mobility (5000-11000 cm²/Vs). This transition is attributed to the milling penetration depth exceeding the STO lattice constant (3.905 Å). Our results provide valuable insights for optimizing STO-based device fabrication, enabling precise control over surface properties while maintaining desired insulating characteristics.

## I. INTRODUCTION

Wide-bandgap semiconductor SrTiO$_3$ (STO) has emerged as a promising material for next-generation electronics, particularly in the field of oxide electronics,[1,2,3,4] due to its exceptional properties of high dielectric constant (100-200),[4,5] high electron mobility,[6,7] almost 100% quantum efficiency of photocatalytic water splitting under ultraviolet light (UV),[8,9] and compatibility as a substrate for complex oxide heterostructures.[2,10,11] These unique characteristics of STO make it an ideal candidate for various electronic and photonic applications, including tunneling barriers,[12] high-mobility channels,[13] high-$k$ capacitors,[4,5] and photocatalytic material.[8,9] To fully harness STO's potential, precise and reliable device processing techniques need to be developed. Among these techniques, ion milling has gained significant attention as a versatile dry etching process applicable to various materials. It is widely used to fabricate numerous electronic components in industry, such as micro electro mechanical systems, acoustic wave filters, tunnel magnetoresistance devices, and other applications, where high precision and control over material removal are required. However, when applied to STO, ion milling induces surface damage that transforms the ion-milled surface from an insulator to a conductor, generating high-mobility carriers at the surface.[14-16] This effect has opened up new possibilities for device processing, utilizing the conductive surface induced by ion milling. Nevertheless, for many applications, minimizing ion milling damage while achieving fine patterning of STO remains a critical challenge. Optimized ion milling processes that mitigate surface damage, therefore, urgently need to be developed to advance the use of STO in electronic devices.

In this study, we systematically investigated the dependence of the surface conduction of STO on ion-milling incident angle ($\theta$) and milling power ($P$). When $\theta$ is 10° or less, the surface remains insulating after milling. In contrast, when $\theta$ exceeds 10°, metallic conduction is induced on the surface. The sheet electron carrier density ($n_s$) and electron mobility ($\mu$) of the high-mobility carriers at the surface state did not show significant changes with variations in $\theta$ and $P$ and were found to be within the ranges of $n_s = 3\times10^{14}$-$8\times10^{14}$ [cm$^{-2}$] and $\mu = 5000$-$11000$ [cm$^2$/Vs]. These results demonstrate that it is possible to control whether the surface state remains insulating or transitions to conducting with high-mobility carriers by adjusting the milling incident angle. This establishes a technique for dry etching STO while preserving the surface's insulating state.

**II. METHODS**

Before ion milling, STO substrates (Crystec GmbH) were annealed in oxygen at 1000°C for 8 hours. The surface morphology after the annealing composed of flat terraces and molecular steps with a height of ~0.4 nm, corresponding to a single unit cell thickness, was determined by atomic force microscopy. Ion milling was carried out by a custom-designed ion beam system built on a commercial platform (10IBE, Hakuto Co., Ltd.). Figure 1(a) shows a schematic diagram of the ion milling process. Ion milling was performed over a range of conditions: incident angle $\theta$ of 5, 10, 15, 30, 60, and 90°; milling powers $P_{150}^{500}$ (Beam voltage $V_B$ = 500 V, Beam current $I_B$ = 150 mA), $P_{106}^{354}$ ($V_B$ = 354 V, $I_B$ = 106 mA), $P_{75}^{250}$ ($V_B$ = 250 V, $I_B$ = 75 mA), $P_{50}^{200}$ ($V_B$ = 200 V, $I_B$ = 50 mA), and $P_{50}^{100}$ ($V_B$ = 100 V, $I_B$ = 50 mA); and milling time of 10 minutes. Here, the superscript and subscript of $P$ represent the values of $V_B$ in units of V and $I_B$ in units of mA, respectively. The angular dependence of the milling rate for $P_{106}^{354}$ estimated with a step gauge is shown in Fig. 1(b). The STO substrate is etched even at a shallow angle of $\theta$ = 10°, and the milling rate is proportional to sin$\theta$. After ion milling, Ag (50 nm)/Al (20 nm) electrodes as ohmic contacts were deposited for the standard Van der Pauw method [Fig. 1(c)]. The $\mu$ and the sheet carrier density $n_s$ are estimated from the Hall-effect measurements.

**III. RESULTS AND DISCUSSIONS**

Figure 2(a) and (b) show the temperature dependence of the longitudinal sheet resistivity $\rho_{xx}$ and the Hall resistivity $\rho_{xy}$ for the STO substrate after ion milling with $\theta$ = 90° and $P$ = $P_{106}^{354}$, as an example. The $\rho_{xx}$ decreases rapidly as temperature decreases, reaching a small sheet resistance of about 1 Ω at 2 K due to the increased electron mobility. The residual resistivity ratio RRR [≡ $\rho_{xx}$(300 K)/$\rho_{xx}$(2 K)] is 2220, which is comparable to the values obtained from bulk STO samples lightly doped with oxygen vacancies or Nb, indicating the high-crystalline quality of the surface conducting layer.[17,18] The $\mu$ and $n_s$ obtained from $\rho_{xx}$ and $\rho_{xy}$ at 2 K are 5723 cm$^2$/Vs and 6.61×10$^{14}$ cm$^{-2}$, respectively. These are typical values for the high-mobility carriers induced on the STO surface by ion milling, as reported previously.[16] These high-mobility carriers come from oxygen vacancies formed during the milling process.[14,16,19]

Figure 3 shows the $\theta$ dependence of $\mu$ and $n_s$ with $P_{106}^{354}$ at 2 and 300 K. The results indicate that metallic conduction occurs when $\theta \geq 15°$, with $\mu$ and $n_s$ values showing minimal variation. This suggests that the electronic properties are largely preserved regardless of incident ion milling angles above 15°. However, at smaller angles of $\theta$ = 5° and 10°, the system remains insulating, with the two-terminal resistance exceeding the measurable limit (> 50 MΩ). At incident ion energies below 1 keV, where nuclear stopping dominates, a well-known empirical formula for ion penetration depth in Å unit is described as[14,20]

$$L = 1.1 \frac{E^{2/3} W_{STO}}{\rho_{STO}(Z_{Ar}^{1/4} + Z_{STO}^{1/4})^2}, \quad (1)$$

where $L$ is the penetration depth, $E$ is the ion energy in eV, $W_{STO}$ is the weighted average of the atomic weight of STO, $\rho_{STO}$ is the density of STO, $Z_{Ar}$ is the atomic number of Ar, and $Z_{STO}$ is the weighted average of the atomic number of STO. On the basis of the relationship of $L = 0.47 E^{2/3}$ derived from eq. (1), the $L$ for $V_B$ of 354 eV is estimated to be 23.5 Å. As summarized in Table 1, the surface of STO remains insulating after milling only when the penetration depth in the vertical direction to the substrate, estimated by $L\sin\theta$, is comparable to or smaller than the STO's lattice constant (3.905 Å). This indicates that by etching only the top surface of STO during milling, the formation of a conducting layer on the surface can be avoided by preventing milling damage to the layers beneath the STO top surface layer. The reason the mobility and carrier density do not significantly change with $L\sin\theta$ at $\theta \geq 15°$, despite the different $L\sin\theta$ values, is thought to be that oxygen vacancies diffuse within the film after milling, resulting in oxygen vacancies being distributed nearly identically.[15,21] This finding demonstrates that the control of the STO surface state through tuning of milling angles significantly influences the material's electronic characteristics, thereby enabling the optimization of advanced microfabrication techniques that rely on tailored surface and interface properties.

According to eq. (1), even if the incident angle is maintained at $\theta = 90°$, reducing $V_B$ can shorten the penetration depth in the vertical direction of the STO substrate. However, to reduce $L$ to 3.905 Å, which is the lattice constant of STO, $V_B$ would have to be lowered to 24 V. At such a low voltage, the ion beam is impossible to stabilize, and even if it were stabilized, the etching rate would be extremely low. In fact, at $P_{50}^{100}$ with $\theta = 90°$, the etching rate in our system is already quite low, at 0.17 nm/min. The Hall-effect measurement results at powers equal to or greater than $P_{50}^{200}$, which maintain a practical etching rate of 1.35 nm/min, with $\theta = 90°$ are shown in Figure 4. Despite a significant reduction in both $\mu$ value at 2 K (by over two orders of magnitude compared to other conditions) and $n_s$ (by more than a quarter compared to other conditions) at $P_{50}^{200}$, metallic conduction persists across all experimental conditions. This suggests that simply adjusting power is insufficient to retain the insulating state of the STO surface after the ion milling. Another method to make the surface layer insulating is to reduce oxygen vacancies by annealing at a high temperature of 500°C in an oxygen atmosphere after milling.[22,23] However, post-processing in such a high-temperature environment often damages other oxide layers included in heterojunctions. For example, ferromagnetic metal $SrRuO_3$,[24] which is a typical metal electrode in oxide electronics, has been reported to experience an increase in resistance due to Ru deficiency and decomposition at temperatures above 300°C.[25,26]

As an example of a device process utilizing the control of the surface state after milling, we fabricated vertical mesa structures that maintain electrical insulation from each other. When fabricating vertical mesa structures, including heterostructures with STO, $Ar^+$ ions mill the STO edges horizontally, which is expected to maintain an insulating state, since the ions do not penetrate the side walls. To verify this, we fabricated a vertical mesa structure schematically shown in Fig. 5 and examined the electrical properties between two vertical mesas. To fabricate the vertical mesa structures, a thin $SrRuO_3$ film was grown as a metal electrode on the STO substrate using machine-learning-assisted molecular beam epitaxy,[27-28)] followed by ion milling with $P_{106}^{354}$ and $\theta = 90°$. The resistance between the two SRO electrodes exceeds the measurable limit (> 50 MΩ), showing that the sidewalls of the vertical mesas remain insulating. These findings suggest that the surface of STO can remain insulating when ions are supplied, in turn, from the cross-sectional direction (parallel to the STO surface), which is primarily used for preparing samples in electron microscopy observations (such as scanning electron microscopy and transmission electron microscopy).

**IV. CONCLUSIONS**

In this study, we have demonstrated that the electrical properties of $SrTiO_3$ (STO) from conducting with high-mobility carriers to insulating can be quantitatively controlled through ion milling techniques. Our findings indicate that the incident angle during ion milling plays a critical role in determining the electrical properties of the surface states. Specifically, an incident angle of 10° or less retains the insulating characteristics, while angles above this threshold induce metallic conduction with minimal angle dependence in electron mobility and carrier density. The ability to fine-tune the surface state of STO opens new avenues for advanced microfabrication techniques, particularly in applications requiring high-precision etching while maintaining insulating properties. Additionally, the successful fabrication of vertical mesa structures without leak currents through the edges further validates the potential for preserving an insulating state even after milling processing.

**CONFLICT OF INTEREST**
The authors have no conflicts of interest to disclose.

**AUTHORS' CONTRIBUTIONS**

**Y. K. Wakabayashi**: Conceptualization (lead); Validation (lead); Investigation (lead); Supervision (lead); Writing – Original Draft (lead); Writing – Review & editing (lead). **Y. Krockenberger, K. Takiguchi**: Investigation (supporting); Writing – Review & editing (supporting). **H. Yamamoto, Y. Taniyasu**: Writing – Review & editing (supporting).

**DATA AVAILABILITY**
The data that support the findings of this study are available from the corresponding author upon reasonable request.

**Figures and Table**

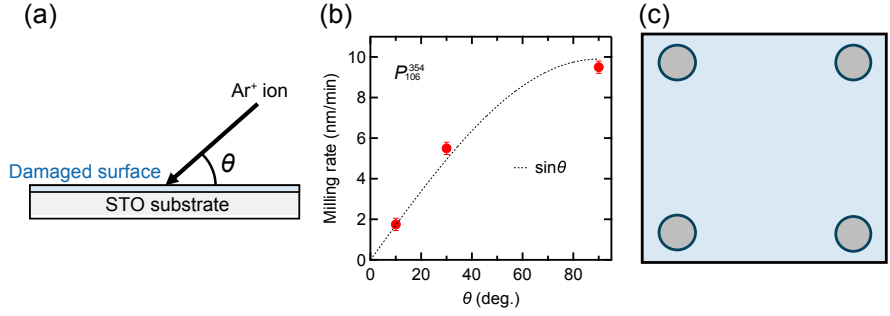

**Fig. 1.** (a) Schematic diagram of the ion milling process for STO substrates. (b) Dependence of the milling rate of STO on the argon-ion incident angle $\theta$, with $P_{106}^{354}$. The dashed curve shows the $\sin\theta$ fitting curve. (c) Schematic diagram of the STO sample for the standard Van der Pauw method with four Ag (50 nm)/Al (20 nm) electrodes (gray circles) on a 4 mm × 4 mm rectangular STO substrate (blue square).

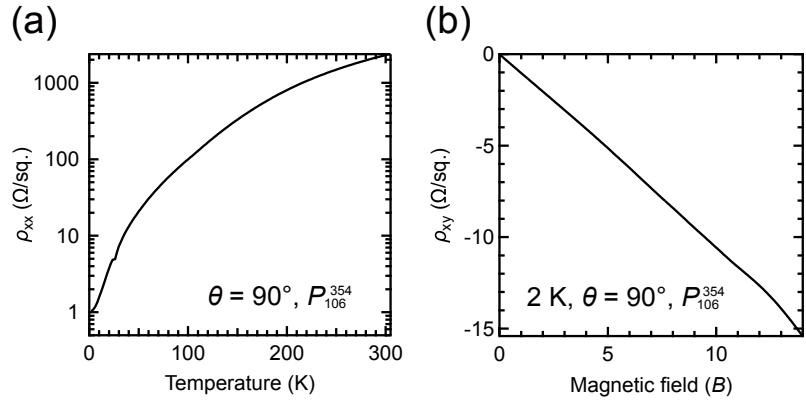

**Fig. 2.** (a) Temperature dependence of the longitudinal resistivity $\rho_{xx}$ for the STO substrate after ion milling with $\theta = 90°$ and $P_{106}^{354}$. (b) Magnetic field $B$ dependence of $\rho_{xy}$ for the STO substrate after ion milling with $\theta = 90°$ and $P_{106}^{354}$.

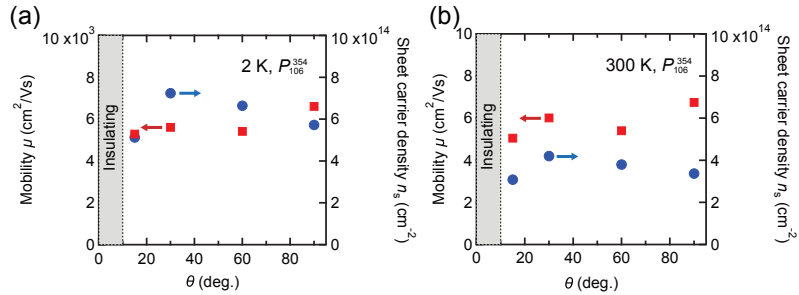

**Fig. 3.** Ion milling angle $\theta$ dependence of $\mu$ and $n_s$ with $P_{106}^{354}$ at (a) 2 and (b) 300 K. At $\theta = 5$ and $10°$, insulating means that the two-terminal resistance exceeds the measurable limit (> 50 MΩ).

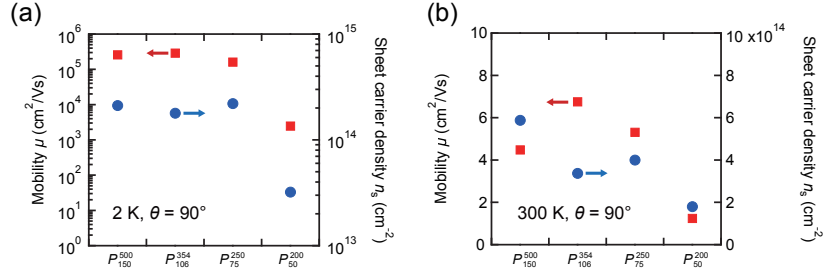

**Fig. 4.** $\mu$ and $n_s$ under the various $P$ conditions with $\theta = 90°$ at (a) 2 and (b) 300 K.

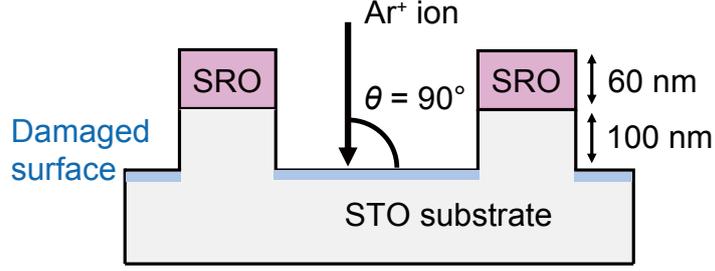

**Fig. 5.** Schematic diagram of the vertical mesa structures. The two mesas have a 1 mm × 2 mm rectangular shape and are separated by a distance of 1.5 mm.

**Table I.** The relationship between the incident angle $\theta$ and penetration depth in the vertical direction to the substrate $L\sin\theta$ at $P_{106}^{354}$, and the surface state after 10 minutes of milling.

| $\theta$ (degree) | $L\sin\theta$ (Å) | Surface state |
|---|---|---|
| 90 | 23.5 | Conducting |
| 60 | 20.4 | Conducting |
| 30 | 11.8 | Conducting |
| 15 | 6.1 | Conducting |
| 10 | 4.1 | Insulating |
| 5 | 2.0 | Insulating |